\documentclass[a4paper,11pt]{article}
\usepackage[top=1 in,bottom=1 in,left=1.0 in,right=1.0 in]{geometry}

\usepackage{array} 
\usepackage{float}

\usepackage{amssymb,amsmath,amsthm}
\usepackage{color}
\usepackage{accents}
\usepackage{texdraw}
\usepackage{eucal}
\usepackage{todonotes}
\usepackage{epic,epsfig}
\usepackage{graphicx}
\usepackage[all]{xy}
\usepackage{tikz}
\usepackage{mathrsfs}
\usepackage[colorlinks,linkcolor=red,linkcolor=blue,linktocpage]{hyperref}
\hypersetup{colorlinks=true, citecolor=blue}

\newtheorem{remark}{Remark}
\numberwithin{equation}{section}
\DeclareMathAccent{\wtilde}{\mathord}{largesymbols}{"65}
\DeclareMathAccent{\what}{\mathord}{largesymbols}{"62}

\def\m@th{\mathsurround=0pt}
\mathchardef\bracell="0365
\def\upbrall{$\m@th\bracell$}
\def\undertilde#1{\mathop{\vtop{\ialign{##\crcr
    $\hfil\displaystyle{#1}\hfil$\crcr
     \noalign
     {\kern1.5pt\nointerlineskip}
     \upbrall\crcr\noalign{\kern1pt
   }}}}\limits}

\newcommand{\wh}{\widehat}
\newcommand{\wt}{\widetilde}

\newcommand{\bblu}{\begin{color}{blue}}
\newcommand{\bred}{\begin{color}{red}}
\newcommand{\ecl}{\end{color}}

\newcommand{\bI}{\boldsymbol{I}}

\newcommand{\bK}{\boldsymbol{K}}

\newcommand{\bM}{\boldsymbol{M}}

\newcommand{\br}{{\boldsymbol{r}}}

\newcommand{\be}{\begin{equation}}
\newcommand{\ee}{\end{equation}}
\newcommand{\bea}{\begin{eqnarray}}
\newcommand{\eea}{\end{eqnarray}}
\newcommand{\bse}{\begin{subequations}}
\newcommand{\ese}{\end{subequations}}
\newcommand{\nn}{\nonumber}

\newcommand{\bc}{\boldsymbol{c}}

\def \h#1{\widehat{#1}}
\def \t#1{\widetilde{#1}}

\begin{document}
\title{On the formulation of the NQC variable}

\author{Leilei Shi$^{1,2}$, ~~ Cheng Zhang$^{1,2}$,
	~~ Da-jun Zhang$^{1,2}$\footnote{Corresponding author.
		Email: djzhang@staff.shu.edu.cn}
	~~\\
	{\small $~^1$Department of Mathematics, Shanghai University, Shanghai 200444,    China} \\
	{\small $^{2}$Newtouch Center for Mathematics of Shanghai University,  Shanghai 200444, China}
	}

\maketitle

\begin{abstract}

The Nijhoff-Quispel-Capel (NQC) equation is a general lattice quadrilateral equation
presented in terms of a function $S(a,b)$ where $a$ and $b$ serve as extra parameters.
It can be viewed as the counterpart of Q3 equation which is the second top equation
in the  Adler-Bobenko-Suris list.
In this paper, we review some known formulations of the NQC variable $S(a,b)$,
such as the Cauchy matrix approach  and the eigenfunction approach via a spectral Wronskian.
We also present a new perspective to formulate $S(a,b)$ from the eigenfunctions of  a Lax pair of the
lattice (non-potential) modified Korteweg-de Vries equation.
A new Dbar problem is introduced and employed in the derivation.

\begin{description}
\item[Keywords:] Nijhoff-Quispel-Capel equation; eigenfunction; Dbar problem
\item[Mathematics Subject Classification:] 37K60
\end{description}
\end{abstract}

%
%

\section{Introduction}\label{sec-1}

The Nijhoff-Quispel-Capel (NQC) equation takes the form
\begin{equation}\label{NQCeq}
 \frac{1-(\alpha +b)\wh{\wt{S}}(a,b)+(\alpha -a)\wh{S}(a,b)}
{1-(\beta +b)\wh{\wt{S}}(a,b)+(\beta -a)\wt{S}(a,b)}
=\frac{1-(\beta +a)\wh{S}(a,b)+(\beta -b)S(a,b)}
{1-(\alpha +a)\wt{S}(a,b)+(\alpha -b)S(a,b)},
\end{equation}
where $S:=S(a,b)$ is a function of $(n,m)\in \mathbb{Z}^2$,
$\alpha$ and $\beta$ are respectively two spacing parameters of $n$- and $m$-direction,
$a$ and $b$ are two extra parameters.
Here conventional notations have been adopted  to denote shifts in discrete independent variables, i.e.
\[S=S_{n,m},~~ \wt S=S_{n+1,m},~~ \wh S=S_{n,m+1},~~ \h{\t S}=S_{n+1,m+1}.\]
We call function $S(a,b)$ the NQC variable for convenience  in this paper.
In continuum limit, the NQC equation \eqref{NQCeq} yields \cite{NC-AAM-1995}
\begin{equation}\label{NQC-c}
S_t=S_{xxx}+3\frac{(S_{xx}+a S_x)(S_{xx}+b S_x)}{1+(a+b)S-2S_{x}},
\end{equation}
which is a generalization of the Schwarzian Korteweg-de Vries (KdV) equation.
The NQC equation was first derived by Nijhoff, Quispel and Capel
in 1983 from the direct linearization approach \cite{NQC-PLA-1983} and bears their names.
It is  general in the lattice KdV type equations.
In fact, in the celebrated Adler-Bobenko-Suris (ABS) list \cite{ABS-CMP-2003} for the quadrilateral equations
that are consistent-around-cube,
Q3$(\delta)$ equation is the equation on the second top of the list:
all other equations (except the top equation Q4) in the ABS list can be obtained as its degenerations
(see, e.g.\cite{NAH-JPA-2009}).
Equation Q3$(\delta=0)$ is nothing but the NQC equation \eqref{NQCeq} up  to a simple transformation;
and moreover, solutions of Q3$(\delta)$ can be obtained as a linear combination
of the different NQC variables $S(\varepsilon_1 a, \varepsilon_2 b)$ where $\varepsilon_i \in \{+1,-1\}$
\cite{AHN-JPA-2008,NAH-JPA-2009}.
Except for the direct linearization approach, the NQC equation \eqref{NQCeq}
together with its solutions can also be formulated
from the Cauchy matrix approach \cite{NAH-JPA-2009}.
In addition, this equation has been bilinearized \cite{ZZ-JNMP-2019}
and the bilinear equations are  the Hirota-Miwa equation in different directions
with reflection symmetries \cite{WZM-PD-2024}.
The NQC variable $S(a,b)$ is characterized by possessing two extra parameters $a$ and $b$.
Apart from \eqref{NQCeq}, such type of equations were also extended to
the lattice Boussinesq case \cite{ZZN-SAPM-2012} and the
lattice Kadomtsev-Petviashvili (KP) case \cite{HJN-book-2016}.

One purpose of the paper is to understand the NQC variable $S(a,b)$
from the perspective of eigenfunctions.
In two recent works \cite{SZZ-arXiv-2025,WZZZ-JPA-2024}, $S(a,b)$ was formulated in terms of
the eigenfunctions of the Lax pair of the lattice potential KdV (lpKdV) equation
(i.e. H1 equation in the ABS list).
The formulation involved   two eigenfunctions with different parameters $a,b$ \cite{WZZZ-JPA-2024},
which was later interpreted in \cite{SZZ-arXiv-2025} as a quasi spectral Wronskian related to a Dbar problem
for the lpKdV equation.
In the present paper, we will show that the NQC variable $S(a,b)$ can be
formulated by a single eigenfunction of a Lax pair (see \eqref{Lax-mkdv} and \eqref{Lax-mkdv2})
of the non-potential lattice modified KdV (lmKdV) equation.

Our another purpose is to understand the master functions $\{S^{(i,j)}\}$ (see Sec.\ref{sec-2})
of the Cauchy matrix approach from the perspective of eigenfunctions as well.
We have shown in \cite{SZZ-arXiv-2025} that all $\{S^{(0,j)}\}$
could be formulated from the eigenfunctions related to the lpKdV equation,
but the link to $\{S^{(i,j)}\}$ for arbitrary $(i,j)$ was still unrevealed.
This gap will be filled in this paper after we formulate the  NQC variable $S(a,b)$ using a single eigenfunction.

The paper is organized as follows.
In Sec.\ref{sec-2} we briefly review how the NQC equation \eqref{NQCeq}
was formulated from the Cauchy matrix approach and
from the Dbar problem of the lpKdV equation.
Then, in Sec.\ref{sec-3} we introduce a new Dbar problem
and a Lax pair related to a lattice non-potential mKdV  equation,
and explain how the NQC equation comes out from the  eigenfunction of the Lax pair.
Sec.\ref{sec-4} is devoted to presenting an explicit
expression for the NQC variable $S(a,b)$.
And as a result, the Cauchy matrix variables  $\{S^{(i,j)}\}$ with arbitrary $(i,j)$
can be formulated from the  single eigenfunction.
Finally, concluding remarks are given in  Sec.\ref{sec-5}.
There are three appendices. 
The first one describes a generalized Cauchy integral formula,
$\delta$ function on complex plane and aa assumption about  the uniqueness of solutions of Dbar equations.
The second one provides some details
of the compatibility results of the Lax pair \eqref{Lax-mkdv2}.
The third one introduces  a new formulation for $S(a,b)$, which is different from ours.

\section{The NQC equation from Cauchy matrix approach and Dbar problem}\label{sec-2}

In what follows  we will introduce how the master functions $\{S^{(i,j)}\}$ and
the NQC variable $S(a,b)$ are defined in the Cauchy matrix approach.
We will also recall the  Dbar problem of the lpKdV equation
and the formulation of the NQC equation.

\subsection{Cauchy matrix approach to the lattice KdV type equations}\label{sec-2-1}

For this part one can refer to \cite{NAH-JPA-2009} or Chapter 9 of \cite{HJN-book-2016} or \cite{ZZ-SAPM-2013} 
for more details.
We start from a Sylvester equation
\begin{equation}
\bM \bK+ \bK \bM = {\br}\bc^T,
\label{rel-MK}
\end{equation}
where
\begin{equation}
\bK=\mathrm{diag}(k_1,k_2,\cdots,k_N),~~
\br=(\rho_1,\rho_2,\cdots,\rho_N)^T, ~~
\bc=(c_1,c_2,\cdots,c_N)^T,
\end{equation}
$\rho_i$ is  the plane wave factor defined as
\begin{equation}
\rho_i=\biggl(\frac{\alpha  +k_i}{\alpha  -k_i}\biggr)^n\biggl(\frac{\beta  + k_i}{\beta -k_i}\biggr)^m \rho^{(0)}_{i},
\end{equation}
$c_i$, $k_i$ and $\rho^{(0)}_{i}$ are constants, $\bM$ is a``dressed'' Cauchy matrix
\begin{equation}
\bM=(M_{i,j})_{N\times N},~~M_{i,j}=\frac{\rho_i c_j}{k_i+k_j}.
\end{equation}
The master functions $\{S^{(i,j)}\}$ are defined as
\begin{equation}
S^{(i,j)}= \bc^T \,\bK^j(\bI+ \bM)^{-1} \bK^i \br, ~~ ~ i,j\in \mathbb{Z},
\label{Sij-1}
\end{equation}
which have a  symmetry property, i.e.  (see \cite{ZZ-SAPM-2013} for a proof)
\begin{equation}
S^{(i,j)}=S^{(j,i)},~~ ~ i,j\in \mathbb{Z}.
\label{Sij-1-sym}
\end{equation}
It can be proved that $\{S^{(i,j)}\}$ obey dynamical recurrence relations \cite{HJN-book-2016,NAH-JPA-2009}:
\begin{subequations}
\begin{align}
& \alpha  \wt{S}^{(i,j)}-\wt{S}^{(i,j+1)}
=\alpha  S^{(i,j)}+S^{(i+1,j)}-\wt{S}^{(i,0)}S^{(0,j)},\label{eq:Sij-dyna-1}\\
& \alpha  S^{(i,j)}+S^{(i,j+1)}
=\alpha  \wt{S}^{(i,j)}-\wt{S}^{(i+1,j)}+S^{(i,0)}\wt{S}^{(0,j)},\label{eq:Sij-dyna-2}\\
& \beta  \wh{S}^{(i,j)}-\wh{S}^{(i,j+1)}
=\beta  S^{(i,j)}+S^{(i+1,j)}-\wh{S}^{(i,0)}S^{(0,j)},\label{eq:Sij-dyna-3}\\
& \beta  S^{(i,j)}+S^{(i,j+1)}
=\beta
\wh{S}^{(i,j)}-\wh{S}^{(i+1,j)}+S^{(i,0)}\wh{S}^{(0,j)}.\label{eq:Sij-dyna-4}
\end{align}
\label{Sij-1-dyna}
\end{subequations}
The NQC variable $S(a,b)$ is defined as
\begin{equation}
S(a,b)=\bc^T (b\bI+\bK)^{-1}(\bI+\bM)^{-1}(a\bI+\bK)^{-1}\br,~~ a,b\in \mathbb{C},
\label{Sab-1}
\end{equation}
which also has a symmetry property (see \cite{ZZ-SAPM-2013} for a proof)
\begin{equation}
S(a,b)=S(b,a)
\label{Sab-sym}
\end{equation}
and obeys the shift relations \cite{HJN-book-2016,NAH-JPA-2009}
\begin{subequations}
\label{Sab-1-re}
\begin{align}
& 1-(\alpha +b)\wt{S}(a,b)+(\alpha -a)S(a,b)=\wt{V}(a)V(b), \label{Sab-1-re-a}\\
& 1-(\beta +b)\wh{S}(a,b)+(\beta -a)S(a,b)=\wh{V}(a)V(b), \label{Sab-1-re-b}
\end{align}
\end{subequations}
where
\begin{eqnarray}
V(a)=1-\bc^T(a \bI+ \bK)^{-1}(\bI+\bM)^{-1}\br=1-\bc^T (\bI+\bM)^{-1}(a\bI+ \bK)^{-1}\br.
\label{Va}
\end{eqnarray}
Note that after exchanging $a$ and $b$ in \eqref{Sab-1-re-a} and making use of the symmetry property \eqref{Sab-sym} 
we have 
\begin{equation}
1-(\alpha +a)\wt{S}(a,b)+(\alpha -b)S(a,b)=\wt{V}(b)V(a). \label{Sab-1-re-aa}
\end{equation}
Then, eliminating $\wt{S}(a,b)$ from \eqref{Sab-1-re-a} and \eqref{Sab-1-re-aa}, we obtain
an alternative expression of $S(a,b)$:
\begin{eqnarray}\label{Sdbv}
    S(a,b)=\frac{1}{a+b}-\frac{1}{a^2-b^2}[(\alpha+a)\widetilde{V}(a)V(b)-(\alpha+b)V(a)\widetilde{V}(b)].
\end{eqnarray}

With the help of the symmetry properties \eqref{Sij-1-sym} and \eqref{Sab-sym},
lattice equations of the KdV type can be obtained as closed forms of the recurrence relations
\eqref{Sij-1-dyna} and \eqref{Sab-1-re}.
One may refer to \cite{NAH-JPA-2009} or Chapter 9 of \cite{HJN-book-2016} for more details.
Below we just list them out:
\begin{itemize}
\item{lpKdV equation  $(u=S^{(0,0)})$:
\begin{equation}
 (\alpha +\beta +u-\widehat{\widetilde{u}})(\alpha -\beta +\h{u}-\t{u})=\alpha ^2-\beta ^2;\label{lpKdV-a}
\end{equation}
}
\item{lattice potential mKdV (lpmKdV) equation $(v=1-S^{(0,-1)})$:
\begin{equation}
\label{lpmKdV}
  \alpha (v\wh{v}-\wt{v}\wh{\wt{v}})=\beta (v\wt{v}-\wh{v}\wh{\wt{v}});
\end{equation}
}
\item{lattice  Schwarzian KdV (lSKdV) equation $(z=S^{(-1,-1)}-\frac{n}{\alpha }-\frac{m}{\beta })$:
\begin{equation}
\label{lSKdV}
\frac{(z-\wt{z})(\wh{z}-\wh{\wt{z}})}{(z-\wh{z})(\wt{z}-\wh{\wt{z}})}=\frac{\beta ^2}{\alpha ^2};
\end{equation}
}
\item{NQC equation, i.e. \eqref{NQCeq}:
\begin{equation}\label{NQC}
 \frac{1-(\alpha +b)\wh{\wt{S}}(a,b)+(\alpha -a)\wh{S}(a,b)}
{1-(\beta +b)\wh{\wt{S}}(a,b)+(\beta -a)\wt{S}(a,b)}
=\frac{1-(\beta +a)\wh{S}(a,b)+(\beta -b)S(a,b)}
{1-(\alpha +a)\wt{S}(a,b)+(\alpha -b)S(a,b)}.
\end{equation}
}
\end{itemize}

\subsection{NQC equation and Dbar problem related to the lpKdV equation}\label{sec-2-2}

For this part one can refer to \cite{SZZ-arXiv-2025} for details and for more references about the Dbar method.

The Dbar problem to characterize  the lpKdV equation \eqref{lpKdV-a} is \cite{SZZ-arXiv-2025}
\begin{subequations}\label{Dbar-1}
\begin{eqnarray}\label{Dp}
    \bar{\partial} \psi\left( p\right) =\psi \left( -p\right) R\left( p\right) , \qquad p\in\mathbb{C},
\end{eqnarray}
with normalization condition
\begin{equation}\label{cnc}
    \psi  ( p ) =1+\psi^{(-1)}p^{-1}+\psi^{(-2)}p^{-2}+\cdots, \qquad \text{as}  \quad  p\rightarrow \infty.
\end{equation}
Here $p$ serves as a spectral parameter, $\bar{p}$ stands for the complex conjugate of $p$,
$\bar{\partial}=\frac{\partial}{\partial \bar{p}}$, and
\begin{equation}\label{dpwf}
    R ( p ) =\left(\frac{\alpha-p}{\alpha+p}\right)^n\left(\frac{\beta-p}{\beta+p}\right)^mR_0 (p),
\end{equation}
\end{subequations}
where $R_0 (p)$ is a distribution in $p$, independent of $n$ and $m$.
With \eqref{Dp} and \eqref{cnc} and making use of the Cauchy-Pompeiu integral formula  
(see eq.\eqref{CP} in Appendix \ref{app-0}), we have
\begin{equation}\label{chi}
    \psi ( p ) = 1+\frac{1}{2\pi {\rm i}}\int_\mathbb{C}
    \frac{\psi ( -\mu) R(\mu) }{\mu -p}{\rm d}\mu \wedge {\rm d}\bar{\mu },
\end{equation}
where the integration is taken over the entire complex plane $\mathbb{C}$.
In principle, solutions of a Dbar problem are not unique,
but in practice, the above formulation will be used to generate solution $\psi(p)$ (see Sec.\ref{sec-4})
and will be also used  to introduce an assumption for the uniqueness of the solutions of the Dbar problem \eqref{Dp}.

\begin{remark}\label{Rem-1}
In Appendix \ref{app-0} we will explain the Cauchy-Pompeiu integral formula and its extension,
definition of $\delta$ function defined in the complex plane,
and the assumption for the uniqueness of the solutions of the Dbar problem \eqref{Dp}.
In this paper, we assume the Dbar problem \eqref{Dp} has only a zero solution $\psi(p)=0$ 
if $\psi(p) \sim 0~ (p\to \infty)$ (see Remark \ref{Rem-3} in Appendix \ref{app-0}). 
In other words, we always assume that the homogeneous integral equation
\begin{equation}\label{chi0}
    \psi(p) = \frac{1}{2\pi \mathrm{i}} \int_{\mathbb{C}}
    \frac{\psi(-\mu)\, R(\mu)}{\mu - p} \, \mathrm{d}\mu \wedge \mathrm{d}\bar{\mu}
\end{equation}
admits only the trivial solution $\psi = 0$ if $\psi(p) \sim 0~ (p\to \infty)$.

\end{remark}

Now we introduce
\begin{subequations}\label{Lax-LM}
\begin{align}
 &   L (p  )\psi(p)=  ( \alpha+p ) \widetilde{\widetilde{\psi}} (p)+h\widetilde{\psi}(p)
 + ( \alpha-p )\psi(p),\label{L}\\
 &     M(p)\psi(p)=(\beta+p)\widehat{\psi}(p)-(\alpha+p)\widetilde{\psi}(p)+g\psi(p),\label{M}
\end{align}
\end{subequations}
where
\begin{subequations}\label{hg}
\begin{align}
&  h=\psi^{(-1)}-\widetilde{\widetilde{\psi}}{}^{(-1)}-2\alpha, \label{h}\\
&  g=\alpha-\beta+\widetilde{\psi}^{(-1)}-\widehat{\psi}^{(-1)}. \label{g}
\end{align}
\end{subequations}
It has been proved that \cite{SZZ-arXiv-2025}
\begin{equation}\label{LM}
L\left(p \right)\psi(p)= 0,~~ M(p)\psi(p)=0
\end{equation}
when $\psi(p)$ is a solution of the Dbar equation \eqref{Dbar-1}.
The compatibility of \eqref{LM}, i.e. $LM=ML$,
gives rise to the lpKdV equation \eqref{lpKdV-a} where $u=-\psi^{(-1)}$.

To construct the NQC equation \eqref{NQCeq} in this Dbar approach,
we consider a generalized spectral Wronskian \cite{SZZ-arXiv-2025} (cf.\cite{JM-JMP-1987})
\begin{equation}\label{D-pq}
    D(p,q)= \begin{vmatrix}
\psi(p) & (\alpha+p)\widetilde{\psi} (p)\\
\psi(q) & (\alpha+q)\widetilde{\psi} (q)
\end{vmatrix},
\end{equation}
where $\psi(q)$ is also a solution of the Dbar problem \eqref{Dbar-1}, i.e.
\begin{eqnarray}\label{Dq}
    \bar{\partial} \psi(q ) =\partial_{\bar{q}}\psi(q)=\psi ( -q ) R ( q) , \qquad q\in\mathbb{C},
\end{eqnarray}
together with the setting \eqref{cnc} and \eqref{dpwf} with $p$ replaced by $q$.
It is also assumed $p^2 \neq q^2$.
Note that $\psi(q)$ is also a solution of the Lax pair \eqref{LM}, i.e.
$L\left(q \right)\psi(q)= 0,~ M(q)\psi(q)=0$.
This indicates that the generalized spectral Wronskian \eqref{D-pq}
can be expressed symmetrically as
\begin{eqnarray}
    D(p,q)=\begin{vmatrix}
\psi(p) & (\beta+p)\widehat{\psi} (p)\\
\psi(q) & (\beta+q)\widehat{\psi} (q)\\
\end{vmatrix}.
\end{eqnarray}
It is the generalized spectral Wronskian $D(p,q)$ that defines the NQC variable $S(p,q)$, via \cite{SZZ-arXiv-2025}
\begin{eqnarray}\label{5.38}
    D(p,q)=(q-p)\big[1-(p+q) S(p,q)\big],
\end{eqnarray}
which satisfies the NQC equation \eqref{NQCeq}. 

Apart from \eqref{cnc}, it was also assumed in \cite{SZZ-arXiv-2025} that
$\psi(p)$ admits an expansion in a neighbourhood of a finite point $c\in \mathbb{C}$, i.e.
\begin{equation}\label{phi-c}
    \psi(p)=\psi_c^{(0)}+\psi_c^{(1)}(p-c)+\psi_c^{(2)}(p-c)^2+\cdots,
\end{equation}
and $\psi_c^{(0)}$ was proved to satisfy
\begin{equation}\label{va}
    (\alpha-c)\psi_c^{(0)}\widehat{\psi}_c^{(0)}-(\beta-c)\psi_c^{(0)}\widetilde{\psi}_c^{(0)}
    +(\beta+c)\widehat{\psi}_c^{(0)}\widehat{\widetilde{\psi}}_c\!\!{}^{(0)}
    -(\alpha+c)\widetilde{\psi}_c^{(0)}\widehat{\widetilde{\psi}}_c\!\!{}^{(0)}=0,
\end{equation}
which is actually the lpmKdV equation \eqref{lpmKdV} with a parameter $c$ after a simple transformation \cite{SZZ-arXiv-2025}.
Noticing that $\psi_c^{(0)}=\psi(p=c)$ which is nothing but the eigenfunction of lpKdV Lax pair
\eqref{LM} with $p=c$, we can conclude that
the lpKdV eigenfunction gives rise to the lpmKdV equation.
For constructing soliton solutions, we introduce $\delta$ function defined on the complex plane $\mathbb{C}$
(see Appendix \ref{app-0} for more details),
which gives rises to \cite{AF-book-2021}
\begin{eqnarray}
    \int_\Omega f(p)\delta(p-p_0){\rm d}p\wedge {\rm d}\bar{p}=f(p_0),
\end{eqnarray}
where the region $\Omega$ contains the point $p_0$ and $f(p)$ is
a sufficiently smooth function in $\Omega$.
After taking
\begin{equation}\label{R0-p}
     R_0 ( p ) = 2\pi \mathrm{i}\sum ^{N}_{j=1}\rho^{(0)}_j \delta \left( p+k_j\right), ~~~~
     k_j, \rho^{(0)}_j \in \mathbb{C},
\end{equation}
it was showed in \cite{SZZ-arXiv-2025} that the following correspondence between the lpKdV eigenfunction $\psi(p)$
and some Cauchy matrix variables (see Sec.\ref{sec-2-1}) holds:
\begin{subequations}\label{Cor}
\begin{align}
& \psi ^{(l)} \to S^{(0,-l-1)},~~~ (l\leq -1),\\
& \psi_0^{(0)}-1 \to S^{(0,-1)},\\
& \psi_0^{(l)}\to S^{(0,-l-1)},~~~ (l \geq 1),\\
& \psi(p) \to V(p). \label{2.32d}
\end{align}
\end{subequations}

\section{Formulating the NQC equation from a single eigenfunction}\label{sec-3}

The formulation of the NQC equation from the lpKdV eigenfunctions
made use of the generalized spectral Wronskian \eqref{D-pq}.
The NQC variable $S(p,q)$ is expressed in a quadratic form composed by
two eigenfunction $\psi(p)$ and $\psi(q)$.
In this section, we will construct the NQC equation
by considering a new Dbar problem which is related to
a Lax pair of the lattice non-potential mKdV equation.
This will enable us to formulate
the NQC equation using just a single eigenfunction.

\subsection{An inhomogeneous Dbar problem}\label{sec-3-1}

Let us consider an inhomogeneous Dbar problem
\begin{subequations}\label{Dbar-2}
\begin{eqnarray}\label{Dp-d}
    \bar{\partial} \phi\left( p\right) =\phi \left( -p\right) R\left( p\right)-2\pi{\rm i}\delta(p+b)
\end{eqnarray}
with a special asymptotic behavior
\begin{equation}
    \phi \left( p\right) \rightarrow 0, ~~\  \ p\rightarrow \infty,
\end{equation}
where $\mathrm{i}^2=-1$, $p$ is spectral parameter, $b$ is a parameter, and $R(p)$ is  defined as \eqref{dpwf}, i.e.
\begin{equation}\label{dpwf-2}
    R ( p ) =\left(\frac{\alpha-p}{\alpha+p}\right)^n\left(\frac{\beta-p}{\beta+p}\right)^mR_0 (p ).
\end{equation}
\end{subequations}
This is different from the usual Dbar problems because of  the additional term, i.e. 
 $-2\pi{\rm i}\delta(p+{b})$.
Through the integral formula \eqref{Cauchy2},
this special Dbar problem is cast to
\begin{equation}\label{Int}
    \phi \left( p\right) =\frac{1}{p+b}+\frac{1}{2\pi\mathrm{ i}}\int_{\mathbb{C}} \frac{\phi( -\mu) R
    ( \mu) }{\mu -p}\mathrm{d}\mu \wedge \mathrm{d}\bar{\mu }
    =\frac{1}{p+b}+F,
\end{equation}
where the integration is taken over the entire complex plane $\mathbb{C}$, and
\begin{equation}\label{F}
F=\frac{1}{2\pi\mathrm{ i}}\int_{\mathbb{C}} \frac{\phi( -\mu) R
    ( \mu) }{\mu -p}\mathrm{d}\mu \wedge \mathrm{d}\bar{\mu }.
\end{equation}
In light of the above expression, we assume $\phi(p)$ has the following
Laurent expansion  at  infinity:
\begin{equation}\label{Laurent-phi}
\phi \left( p\right) =\sum_{j=1}^{+\infty}(-b)^{j-1}p^{-j}+
\sum_{j=1}^{+\infty} \phi^{(-j)} p^{-j},
\end{equation}
where the two  series correspond to the two terms on the right-hand side of \eqref{Int}, respectively.

With regard to the uniqueness of the solutions of the Dbar problem \eqref{Dbar-2},
we address it in the following remark.

\begin{remark}\label{Rem-2}
Recalling Remark \ref{Rem-3} in Appendix \ref{app-0} where the uniqueness of solutions of the Dbar problem \eqref{Dp}
is addressed,
for \eqref{Dp-d}, in case that we only consider those solutions that are determined from the
 integral formula \eqref{Int},
we assume the homogeneous part of \eqref{Dp-d}, i.e.
\begin{eqnarray}\label{Dp-dh}
    \bar{\partial} \phi\left( p\right) =\phi \left( -p\right) R\left( p\right),
\end{eqnarray}
admits only zero solution $\phi(p)=0$ if $\mathrm{Nor}\,\phi(p)=0$.
Here $\mathrm{Nor}\,\phi(p)$ denotes the polynomial part of the expansion of $\phi(p)$ at infinity,
see \eqref{Nor}. 
We also assume the expansion form \eqref{Laurent-phi} admits
interchangeability with discrete shifts,
i.e.
\begin{equation}\label{Laurent-phi-Tis}
T^i_nT^s_m\phi \left( p\right) =\sum_{j=1}^{+\infty}(-b)^{j-1}p^{-j}+
\sum_{j=1}^{+\infty} T^i_nT^s_m\phi^{(-j)} p^{-j},~~~ i,s\in \mathbb{Z},
\end{equation}
where for a function $S_{n,m}$ defined on $\mathbb{Z}^2$, shift operators $T_n$ and $T_m$ are defined as
\begin{equation}\label{TnTm}
T^i_nT^s_m S_{n,m}=S_{n+i,m+s}.
\end{equation}
\end{remark}

We will see in  Sec.\ref{sec-4} that these assumptions are true at least for soliton solutions.

\subsection{Lattice non-potential mKdV equation}\label{sec-3-2}

Related to the Dbar problem \eqref{Dbar-2}, we consider the following Lax pair,
of which the eigenfunction $\phi(p)$ is a solution of the  Dbar problem \eqref{Dbar-2}:
\begin{subequations}\label{Lax-mkdv}
\begin{equation}\label{LM-mkdv}
 L\left(p \right)\phi(p)=0,~~  M\left(p \right)\phi(p)=0,
\end{equation}
where
\begin{align}
 &   L (p)\phi(p)=  ( \alpha+p) w\widetilde{\widetilde{\phi}} (p)+[-\alpha(1+w)+b(w-1)]\widetilde{\phi}(p)
 + ( \alpha-p )\phi(p), \label{3.6b} \\
& M(p)\phi(p)=- (\beta+p)\widehat{\phi}(p)+ (\alpha+p) s\widetilde{\phi}(p)
-[\alpha  s-\beta -  b (s-1)]\phi(p),\label{3.6c}
\end{align}
$b$ is a parameter, and
\begin{align}
&    w=\frac{1+\phi^{(-1)}}{1+\widetilde{\widetilde{\phi}}{}^{(-1)}}, \label{h-m}\\
&  s=\frac{1+\widehat{\phi}^{(-1)}}{1+\widetilde{\phi}^{(-1)}}, \label{g-m}
\end{align}
\end{subequations}
which are subject to the Laurent expansion of $\phi(p)$, i.e. \eqref{Laurent-phi}.

In light of the expression \eqref{Laurent-phi}
and the above setting for $w$, it follows that
\begin{align}
    L\left(p \right)\phi(p)&\sim ~w\left (1+\widetilde{\widetilde{\phi}}{}^{(-1)}\right )
    -(1+\phi^{(-1)})+O(p^{-1})\nonumber\\
    &\sim ~0+O(p^{-1}), ~~~ (p \sim \infty),
\end{align}
and hence we have 
\begin{eqnarray}
   \mathrm{Nor}\, (L\left(p \right)\phi(p)) =0.
\end{eqnarray}
Moreover,
\begin{align}
\bar{\partial}(L(p)\phi(p))=&
\left( \alpha+p\right) w\widetilde{\widetilde{\phi}} (-p)\widetilde{\widetilde{R}}(p)
+[-\alpha(1+w)+b(w-1)]\widetilde{\phi}(-p)\widetilde{R}(p) \nonumber\\
& + ( \alpha-p )\phi(-p)R(p)-2\pi{\rm i}[(\alpha+p)w-\alpha(1+w)+b(w-1)+\alpha-p]\delta(p+b) \nonumber \\
=& (L\left(-p \right)\phi(-p))\widetilde{ R} ( p )-2\pi{\rm i}(w-1)(p+b)\delta(p+b) \nonumber \\
=& (L\left(-p \right)\phi(-p))\widetilde{ R} ( p ).
\end{align}
This means $L (p )\phi(p)$ with a down-tilde shift satisfies the homogeneous Dbar problem \eqref{Dp-dh}
and then it vanishes in light of the assumption in Remark \ref{Rem-2}.
Along the same line we can find
\begin{eqnarray}
   \mathrm{Nor}\, (M\left(p \right)\phi(p)) =0
\end{eqnarray}
and
\begin{eqnarray}
     \bar{\partial}(M (p)\phi(p))=(M (-p)\phi(-p)) R ( p ),
\end{eqnarray}
which yields $M (p)\phi(p)=0$.
Thus we have got the Lax pair \eqref{Lax-mkdv} where $\phi(p)$ is a solution of the Dbar problem \eqref{Dbar-2}.

Through the gauge transformation
\begin{eqnarray}
    \varphi(p)=(-\alpha)^n(-\beta)^m(\alpha+p)^n(\beta+p)^m\phi(p),
\end{eqnarray}
the Lax pair \eqref{LM-mkdv} together with \eqref{3.6b} and \eqref{3.6c} gives rise to
\begin{subequations}\label{Lax-mkdv2}
\begin{align}
& w\widetilde{\widetilde{\varphi}} (p)
 +\big[\alpha^2(1+w)-\alpha b (w-1)\big]\widetilde{\varphi}(p)+\alpha^2(\alpha^2-p^2)\varphi(p)=0,\\
&  \alpha  \widehat{\varphi}(p)- \beta  s\widetilde{\varphi}(p)-\alpha\beta[(\alpha s-\beta)- b(s-1)]\varphi(p)=0.
\end{align}
\end{subequations}
The above Lax pair with $b=0$ has been given before (\cite{HJN-book-2016}, Eq.(3.91)).
It (with $b=0$) was also constructed as Darboux transformations of the mKdV spectral problem
$\psi_{xx}+v\psi_x+\lambda \psi=0$ in \cite{ZPZ-TMP-2020}.
Keeping the $b$ terms in \eqref{Lax-mkdv2}, from their compatibility we get\footnote{The derivation
with details can be found in Appendix \ref{app-1}.}
\begin{subequations}\label{CCC}
\begin{align}
& \widetilde{s}\, [-b-\alpha+(\beta+b)s ]=w[\beta-b-(\alpha-b)s], \label{CC1}\\
& s\,\widetilde{w}=\widetilde{\widetilde{s}}\,\widehat{w}.\label{CC2}
\end{align}
\end{subequations}
Eliminating $w$ we obtain  a multi-linear quad-lattice equation in $s$, that is
\begin{eqnarray}\label{mkdv}
\frac{\big[\beta-b-(\alpha-b)\widehat{s}\big] \big[-b-\alpha+(\beta+b)\widetilde{s}\,\big]}
{\big[\beta-b-(\alpha-b)\widetilde{s}\big] \big[-b-\alpha+(\beta+b)\widehat{s}\,\big]}
=\frac{\wh{\wt{s}}}{s}.
\end{eqnarray}
This gives the non-potential lattice mKdV equation with a parameter $b$ (cf.\cite{HJN-book-2016}, Eq.(3.90)).

The lpmKdV equation \eqref{lpmKdV} can be obtained from the eigenfunction $\phi(p)$.
In fact, the formulations in \eqref{h-m} and \eqref{g-m} for $w$ and $s$
are the results of \eqref{Laurent-phi}.
If inserting  \eqref{Laurent-phi} into the the Lax pair \eqref{LM-mkdv}  with \eqref{3.6b} and \eqref{3.6c},
one can see that \eqref{h-m} and \eqref{g-m} are respectively obtained from the leading terms
of the two equations in \eqref{LM-mkdv}.
Back to the compatibility result \eqref{CCC}, the second equation \eqref{CC2} is satisfied
in light of \eqref{h-m} and \eqref{g-m},
while the first equation \eqref{CC1} gives rise to the lpmKdV equation with a parameter $b$ (cf.\eqref{va})
\begin{equation}\label{Eq_w}
    (\beta-b)v\widetilde{v}-(\alpha-b)v\widehat{v}
    =(\beta+b)\widehat{v}\widehat{\widetilde{v}}-(\alpha+b)\widetilde{v}\widehat{\widetilde{v}}
\end{equation}
where
\begin{equation}
    v=1+\phi^{(-1)}.
\end{equation}
Note that here $v$ is related to the parameter $b$, i.e. $v=v(b)$.
Then, $s=\h{v}/\t{v}$ is the transformation to connect \eqref{Eq_w} and
its non-potential form \eqref{mkdv}.

\subsection{The NQC equation}\label{sec-3-3}

To achieve the NQC equation, 
we assume  $\phi(p)$ is analytic in a neighbourhood of a finite point $a \in \mathbb{C}$
$ (a \neq -b)$, and then it can be expanded in certain neighborhood of $a$.
Corresponding to the expression \eqref{Int}, we present the expansion in the following form:
\begin{equation}\label{exp-c}
    \phi (p) =-\sum_{j=0}^{+\infty}\frac{(p-a)^j}{(-b-a)^{j+1}}
    +\sum_{j=0}^{+\infty}\phi_a^{(j)}(p-a)^j,
\end{equation}
where $\{\phi_a^{(j)}\}$ stand for coefficients.
Here we assume
\begin{equation}\label{3.21}
T^i_nT^s_m\phi \left( p\right) =-\sum_{j=0}^{+\infty}\frac{(p-a)^j}{(-b-a)^{j+1}}+
\sum_{j=1}^{+\infty} T^i_nT^s_m\phi_a^{(j)}(p-a)^j,~~~ i,s\in \mathbb{Z},
\end{equation}
and note that such interchangeability is valid at least for soliton solutions, see   Sec.\ref{sec-4}.  
Note also that corresponding to the expression \eqref{Int}, here we have
\begin{equation}\label{Int-F}
F= \frac{1}{2\pi\mathrm{ i}}\int_{\mathbb{C}} \frac{\phi( -\mu ) R
    ( \mu) }{\mu -p}\mathrm{d}\mu \wedge \mathrm{d}\bar{\mu }
    =\sum_{j=0}^{+\infty}\phi_a^{(j)}(p-a)^j.
\end{equation}
Substituting the expansion \eqref{exp-c} into  Lax pair \eqref{LM-mkdv}, we get
\begin{subequations}\label{3.18}
\begin{align}
&  \big(1-(\alpha-b)\widetilde{\phi}_a^{(0)} +(\alpha+a)\widetilde{\widetilde{\phi}}_a\!\!{}^{(0)}\big) w
-\big(1-(\alpha-a)\phi_a^{(0)}+(\alpha+b)\widetilde{\phi}_a^{(0)}\big)=0, \label{rr5}\\
& (\alpha+a)w\widetilde{\widetilde{\phi}}_a\!\!{}^{(i)}
+w\widetilde{\widetilde{\phi}}_a\!\!{}^{(i-1)}
-\big(\alpha(1+w)-b(w-1)\big)\widetilde{\phi}_a^{(i)}
+(\alpha-a)\phi_a^{(i)}-\phi_a^{(i-1)}= 0, ~~\mathrm{for} ~ i \geq 1,  \label{rr6}
\end{align}
\end{subequations}
and
\begin{subequations}\label{3.199}
\begin{align}
& -\big((1-(\beta-b)\phi_a^{(0)}+(\beta+a)\widehat{\phi}_a^{(0)})\big)
+s\big((1-(\alpha-b)\phi^{(0)}+(\alpha+a)\widetilde{\phi}^{(0)}\big)=0, \label{rr7}\\
& - (\beta+a)\widehat{\phi}_a^{(i)}- \widehat{\phi}_a^{(i-1)}
-(\alpha+a)  s\widetilde{\phi}_a^{(i)}+  s\widetilde{\phi}_a^{(i-1)}
-\big(\alpha  s-\beta -b ( s-1)\big)\phi_a^{(i)}=0,  ~~ \mathrm{for}~ i \geq 1. \label{rr8}
\end{align}
\end{subequations}
Now we introduce a new variable
\begin{eqnarray}\label{S}
    S:=S(a,b)=-\phi_a^{(0)}
\end{eqnarray}
and then express $w$ and $s$  from \eqref{rr5} and \eqref{rr7}:
\begin{equation}\label{hg-S}
w =\frac{1+(\alpha-a)S -(\alpha+b)\widetilde{S} }{1+(\alpha-b)\widetilde{S} -(\alpha+a)\widetilde{\widetilde{S}} },
~~~
s= \frac{1+(\beta-b)S -(\beta+a)\widehat{S} }{1+(\alpha-b)S -(\alpha+a)\widetilde{S} }.
\end{equation}
Then, substituting them into the compatibility equation \eqref{CC1} yields
 a difference equation in $S $, i.e.
\begin{eqnarray}
\big(1+(\alpha-b)S -(\alpha+a)\widetilde{S} \big)\big(1+(\alpha-a)\widehat{S}
-(\alpha+b)\widehat{\widetilde{S}} \big) \nn \\
=\big(1+(\beta-b)S -(\beta+a)\widehat{S} \big)\big(1+(\beta-a)\widetilde{S}
-(\beta+b)\widehat{\widetilde{S}} \big),
\label{NQC-hg}
\end{eqnarray}
which is nothing but the NQC equation \eqref{NQCeq}. Combining the expressions \eqref{exp-c} and \eqref{S}, we see that
\begin{equation}\label{S-phi}
S =\frac{1}{b+a}-\phi(a)
\end{equation}
where $\phi(a)$ is the eigenfunction of the Lax pair \eqref{Lax-mkdv} when $p=a$.
Thus, we  conclude that the NQC variable $S(a,b)$
can be characterized by a single eigenfunction of the Lax pair of the lattice mKdV equation.

In addition, eliminating $w$ and $s$ from \eqref{h-m}, \eqref{g-m} and \eqref{hg-S},
we obtain two Miura type links
\begin{equation}
 \frac{v}{\widetilde{\widetilde{v}}}
 =\frac{1+(\alpha-a)S-(\alpha+b)\widetilde{S}}{1+(\alpha-b)\widetilde{S}-(\alpha+a)\widetilde{\widetilde{S}}}, \  \ ~~ \frac{\widehat{v}}{\widetilde{v}}
 =\frac{1+(\beta-b)S-(\beta+a)\widehat{S}}{1+(\alpha-b)S-(\alpha+a)\widetilde{S}},
\end{equation}
which connect the lpmKdV equation \eqref{Eq_w} and the NQC equation \eqref{NQC-hg}.

The definition \eqref{S-phi} for $S(a,b)$ does not allow the degeneration $S(-b,b)$.
However, we can obtain an equation of $S(-b,b)$ by the following trick.
Consider the expression \eqref{Int} for $\phi(p)$ where we assume
the term $F$
is analytic at $p=-b$.
In other words, it is the term $1/(p+b)$ to contribute  a simple pole of $\phi(p)$ at certain neighbourhood of $-b$.
Thus, at this neighbourhood  we have (cf. Eq.\eqref{Int-F})
\begin{equation}\label{Int-F-b}
F= \frac{1}{2\pi\mathrm{ i}}\int_{\mathbb{C}} \frac{\phi( -\mu) R
    ( \mu) }{\mu -p}\mathrm{d}\mu \wedge \mathrm{d}\bar{\mu }
    =\sum_{j=0}^{+\infty}\phi_{-b}^{(j)}(p+b)^j,
\end{equation}
which then brings the following:
\begin{equation}\label{exp-b}
    \phi \left( p\right) =\frac{1}{p-b}+\sum_{j=0}^{+\infty}\phi_{-b}^{(j)}(p+b)^j.
\end{equation}
Substituting it into the Lax pair \eqref{Lax-mkdv2}, we can get
equations analogue to \eqref{3.18} and \eqref{3.199}.
Then, after a similar process as deriving the NQC equation \eqref{NQC-hg},
we can get an equation for $\phi_{-b}^{(0)}$:
\begin{eqnarray}
\big(1-(\alpha+b)\phi_{-b}^{(0)}+(\alpha+b)\widetilde{\phi}_{-b}^{(0)}\big)
\big(1-(\alpha-b)\widehat{\phi}_{-b}^{(0)}+(\alpha-b)\wh{\wt{\phi}}_{-b}\!\!{}^{(0)}\big) \nn \\
=\big(1-(\beta+b)\phi_{-b}^{(0)}+(\beta+b)\widehat{\phi}_{-b}^{(0)}\big)
\big(1-(\beta-b)\widetilde{\phi}_{-b}^{(0)}+(\beta-b)\wh{\wt{\phi}}_{-b}\!\!{}^{(0)}\big),
\label{NQC-bb}
\end{eqnarray}
which is a degeneration of the NQC equation \eqref{NQC-hg} for $a=-b$.
Thus, we may introduce $S (-b,b)$ by
\begin{equation}
S(-b,b)=-\phi_{-b}^{(0)},
\end{equation}
where $\phi_{-b}^{(0)}$ is defined by \eqref{Int-F-b}.
In particular,
\begin{equation}
z=S(0,0)-\frac{n}{\alpha}-\frac{m}{\beta}
=-\phi_0^{(0)}-\frac{n}{\alpha}-\frac{m}{\beta}
\end{equation}
satisfies the lSKdV equation \eqref{lSKdV}, i.e.
\begin{equation}\label{3.30}
\frac{\beta^2}{\alpha^2}=\frac{(\widehat{\widetilde{z}}-\widehat{z})(\widetilde{z}-z)}
{(\widehat{\widetilde{z}}-\widetilde{z})(\widehat{z}-z)}.
\end{equation}

\section{Soliton solutions}\label{sec-4}

To obtain exact form of  solutions, we need to study $\phi(p)$ and
\begin{subequations}
\begin{align}
\phi ^{(l)}&= -\frac{1}{2\pi {\rm i}}
      \int_\mathbb{C} \frac{\phi ( -\mu ) R ( \mu )}{\mu^{l+1}} \mathrm{d}\mu \wedge \mathrm{d}\bar{\mu },
      \quad \text{for} \quad l\leq-1, \label{6.3}\\
\phi_a^{(l)}&=\frac{1}{2\pi {\rm i}}
    \int _{\mathbb{C}}\frac{\phi( -\mu ) R( \mu )}{ (\mu-a)^{l+1} } \mathrm{d}\mu \wedge \mathrm{d}\bar{\mu}.
    \quad \text{for} \quad l\geq 0. \label{6.5}
\end{align}
\end{subequations}
For constructing  $N$-soliton solutions, we choose $R_0(p)$ in \eqref{dpwf-2} as the following:
\begin{equation}\label{R0-p}
     R_0 (p) = 2\pi {\rm i}\sum ^{N}_{j=1}\rho^{(0)}_j \delta \left( p+k_j\right), ~~~~
     k_j, \rho^{(0)}_j \in \mathbb{C},
\end{equation}
with $k_j\neq -b, k_j+k_i\neq 0$ and $k_j\neq k_i\neq0, \forall i,j\in \mathbb{N}$.
Substituting \eqref{dpwf-2} with the above $R_0(p)$ into Eq.\eqref{Int}, we have
\begin{align}
    \phi  ( p )  &=\frac{1}{p+b}+\frac{1}{2\pi \mathrm{i}}\int_{\mathbb{C}} \frac{\phi ( -\mu ) R( \mu) }{\mu -p}
    \mathrm{d}\mu \wedge \mathrm{d}\bar{\mu }\nn  \\
    &= \frac{1}{p+b}+\int_{\mathbb{C}} \frac{\phi (-\mu)
    \left(\frac{\alpha-\mu}{\alpha+\mu}\right)^n
    \left(\frac{\beta-\mu}{\beta+\mu}\right)^m
    \sum ^{N}_{j=1}\rho^{(0)}_j\delta (\mu+k_{j})}{\mu-p}\mathrm{d}\mu \wedge \mathrm{d}\bar{\mu} \nn \\
    &= \frac{1}{p+b}-\sum ^{N}_{j=1}\frac{\phi (k_j)\rho_j}{p+k_j },
    \label{4.33}
\end{align}
where
\begin{equation}\label{pwf-rho}
\rho_j=\left(\frac{\alpha+k_j}{\alpha-k_j}\right)^n\left(\frac{\beta+k_j}{\beta-k_j}\right)^m \rho^{(0)}_j.
\end{equation}
Taking $p=k_1, k_2, \cdots,  k_N$ in the above equation, we have a set of equations
\begin{eqnarray}
    \phi ( k_{i})+\sum ^{N}_{j=1}\frac{\phi(k_j)\rho_j}{k_j+k_i }= \frac{1}{k_i+b},
    ~~~ i=1,2,\cdots, N,
\end{eqnarray}
which can be rewritten in matrix form:
\begin{eqnarray}
    (\bI+\bM)\Phi=(\bK+b\bI)^{-1}\bc
\end{eqnarray}
where $\bI$ is the $N$-th order identity matrix, $\bc=(1,1,\cdots, 1)^T$ is a $N$-th order column vector,
\begin{eqnarray}
    \bM=\left(M_{i,j}\right)_{N\times N},
    ~~ M_{i,j}=\frac{\rho_j}{k_j+k_i },
    ~~ \Phi=\left(\begin{array}{c}
    \phi(k_1)   \\
     \phi(k_2)  \\
    \vdots \\
     \phi(k_N)
\end{array}\right),
~~ \bK=\mathrm{diag}\{k_1, k_2,\cdots, k_N\}.
\end{eqnarray}
Thus we have
\begin{equation}\label{Phi}
\Phi= (\bI+\bM)^{-1}(\bK+b\bI)^{-1}\bc,
\end{equation}
and it also follows from \eqref{4.33} that
\begin{equation}\label{phi-p}
\phi(p)=\frac{1}{p+b}-\br^T(\bK+p \bI)^{-1}\Phi
= \frac{1}{p+b}-\br^T(\bK+p \bI)^{-1}(\bI+\bM)^{-1}(\bK+b\bI)^{-1}\bc,
\end{equation}
where
\begin{equation}
    \br =( \rho_1, \rho_2,\cdots, \rho_N)^T.
\end{equation}
Note that by making use of formulae of Cauchy matrix, 
one can  show that $|\bI+\bM|\neq 0$ when assuming 
$\alpha>\beta>k_{N}>k_{N-1}> \cdots > k_1 >0$ 
(see, e.g. \cite{HJN-book-2016}, page 265).
Next, let us consider $\phi_a^{\left(l\right)}$ given in \eqref{6.5}.
For it we have
\begin{align}\label{phi(-1)}
\phi_a^{\left(l\right)}&=\sum_{j=1}^N\frac{\phi\left(k_j\right)\rho_j}{ (-k_j-a)^{l+1}}
=(-1)^{l+1}\br^T (\bK+a\bI)^{-(l+1)} \left(\bI+\bM\right)^{-1}(\bK+b\bI)^{-1}\bc.
\end{align}
In a similar way, from \eqref{6.3} we  obtain
\begin{equation}\label{phi-l}
\phi ^{(l)}=(-1)^l\br^T \bK ^{-(l+1)} \left(\bI+\bM\right)^{-1}(\bK+b\bI)^{-1}\bc,~~~ (l\leq -1).
\end{equation}

Now we can write out some explicit formulae for solutions.
For the lpmKdV equation \eqref{Eq_w} with a parameter $b$, we have
\begin{eqnarray}\label{vb}
  v=v(b)=1+\phi^{(-1)}=1-\br^T  \left(\bI+\bM\right)^{-1}(\bK+b\bI)^{-1}\bc.
\end{eqnarray}
For the solution of the NQC equation \eqref{NQC-hg}, we have  (see \eqref{S-phi})
\begin{eqnarray}\label{Sab}
    S(a,b)=-\phi(a)+\frac{1}{a+b}= \br^T (\bK+a\bI)^{-1} \left(\bI+\bM\right)^{-1}(\bK+b\bI)^{-1}\bc.
\end{eqnarray}
For the lSKdV equation \eqref{3.30}, there is
\begin{equation}
z= -\frac{n}{\alpha}-\frac{m}{\beta}
+\br^T  \bK^{-1} \left(\bI+\bM\right)^{-1}\bK^{-1}\bc.
\end{equation}

Now we can make a comparison with the Cauchy matrix variables displayed in Sec.\ref{sec-2-1}
and our above expressions.
The Cauchy matrix variables $S(a,b)$ and $V(a)$ are defined in \eqref{Sab-1} and \eqref{Va}.
Comparing them with \eqref{Sab} and \eqref{vb}, they are same (after we replace $b$ with $a$ in \eqref{vb}).
To reveal more links, let us back to the eigenfunction $\phi(p)$ in \eqref{Int} and \eqref{F},
where   $F$ is
\begin{equation}\label{Int-F-pb}
F(p,b)=: \frac{1}{2\pi\mathrm{ i}}\int_{\mathbb{C}} \frac{\phi( -\mu) R
    ( \mu) }{\mu -p}\mathrm{d}\mu \wedge \mathrm{d}\bar{\mu } 
\end{equation}
and the integration is taken over the entire complex plane $\mathbb{C}$.
Here we highlight the variable $p$ and parameter $b$.
From \eqref{Int-F} and \eqref{S-phi}, one can identify that
\begin{equation}
S(a,b)=F(a,b)=\frac{1}{2\pi\mathrm{ i}}\int_{\mathbb{C}} \frac{\phi( -\mu) R
    ( \mu) }{\mu -a}\mathrm{d}\mu \wedge \mathrm{d}\bar{\mu }
    =-\phi(a)+\frac{1}{a+b},
\end{equation}
where $\phi(p)$ is a solution of the Dbar problem \eqref{Dbar-2} as well as the eigenfunction
of the Lax pair \eqref{Lax-mkdv}.
Moreover, for any such eigenfunction $\phi(p)$,
we can expand  $\phi(a)-\frac{1}{a+b}$ in terms of $(a,b)$
in the neighbourhood of $(0,0)$, $(\infty,0)$, $(0,\infty)$ and $(\infty,\infty)$, respectively,
and we get
\begin{subequations}\label{Sij-phi}
\begin{align}
\phi(a)-\frac{1}{a+b}=-S(a,b)&
=\sum^{i=-1}_{-\infty}\sum^{j=-1}_{-\infty}(-1)^{i-j-1}\phi^{(i,j)}a^{-i-1}b^{-j-1},\\
\phi(a)-\frac{1}{a+b}=-S(a,b)&
= \sum^{\infty}_{i=0}\sum^{j=-1}_{-\infty}(-1)^{i-j}\phi^{(i,j)}a^{-i-1}b^{-j-1},   \\
\phi(a)-\frac{1}{a+b}=-S(a,b)&
= \sum^{i=-1}_{-\infty}\sum^{\infty}_{j=0}(-1)^{i-j}\phi^{(i,j)}a^{-i-1}b^{-j-1},    \\
\phi(a)-\frac{1}{a+b}=-S(a,b)&
= \sum^{\infty}_{i=0}\sum^{\infty}_{j=0}(-1)^{i-j-1}\phi^{(i,j)}a^{-i-1}b^{-j-1},
\end{align}
\end{subequations}
where the coefficients give rise to
\begin{eqnarray}
    \phi^{(i,j)}=\br^T \bK^{i}(\bI+ \bM)^{-1}\bK^{j}\bc, ~~~ (i,j \in \mathbb{Z}).
\end{eqnarray}
This indicates how the Cauchy matrix variables $\{S^{(i,j)}\}$ are defined by
the eigenfunction $\phi(p)$ of the Lax pair \eqref{Lax-mkdv}.

\section{Concluding remarks}\label{sec-5}

In this paper we have provided a new formulation for the NQC equation.
It made use of the Lax pair, i.e. \eqref{Lax-mkdv},
of  the lattice non-potential mKdV equation with a parameter $b$,
rather than the Lax pair of the lpKdV equation.
As a result, we can characterize the NQC variable $S(a,b)$ by using the single eigenfunction of
the Lax pair \eqref{Lax-mkdv}.
As we have shown in Sec.\ref{sec-4},  an explicit expression for $S(a,b)$
can be derived from  solving the integral equation \eqref{4.33}.
The formula, i.e. \eqref{Sab},  is the same as the expression in the Cauchy matrix approach.
This also indicates that if we make a binary expansion in terms of $(a,b)$,
then we can obtain all  $\{S^{(i,j)}\}$ from the expansion of $S(a,b)$
(as well as from the eigenfunction $\phi(p)$),
which has been shown in the previous section.
In addition, $V(p)$ can also be formulated by the $\phi(p)$ through expansion \eqref{phi-pp}
(also see \eqref{Vp}).
Thus, we can recover  all the Cauchy matrix variables including $S^{(i,j)}$, $V(p)$ and $S(a,b)$
from the single eigenfunction $\phi(p)$.

Recall our previous paper \cite{SZZ-arXiv-2025}, where the equation characterized by
the single eigenfunction of the lpKdV Lax pair is the lpmKdV equation with a
parameter $c$,  (see \eqref{va} of this paper or Eq.(4.6) in \cite{SZZ-arXiv-2025}).
Here by single eigenfunction characterization we mean
the equation is formulated by just a single eigenfunction, not by their combinations.\footnote{
In Sec.\ref{sec-2-2}, the lpmKdV equation \eqref{va} is formulated by single $\psi(p)|_{p=c}$,
while the NQC variable $S(p,q)$ is formulated by $D(p,q)$ which is a combination of $\psi(p), \psi(q)$
and their shifts, cf.\eqref{va} and \eqref{5.38}.}
Now we can say that the single eigenfunction of the lattice (non-potential) mKdV Lax pair
can be used to characterize the NQC equation.
Thus we have got a hierarchy from the lpKdV and its Lax pair, whose single eigenfunction
gives rise to the lpmKdV equation; and then from the lattice (non-potential) mKdV Lax pair,
its single eigenfunction gives rise to the NQC equation.
Note that in \cite{ZZZ-SHU-2022} a hierarchy from the lpKdV to lpmKdV and to NQC equation
has been made by means of gauge transformations of their Lax pairs.
It is also stated in \cite{ZZZ-SHU-2022} that the gauge transformation of the NQC Lax pair
does not lead to any new lattice equation in terms of eigenfunctions.
However, we will show in Appendix \ref{app-2} the formulations for the NQC variable $S(a,b)$
in  \cite{ZZZ-SHU-2022} and in our paper are completely different.
In the future we will look at what eigenfunction equations come from the NQC equation.
In other words, we will check the Lax pair of the NQC equation and
maybe a new Dbar problem should be introduced.

\vskip 20pt
\subsection*{Data availability}

No data was used for the research described in the article.

\subsection*{Conflict of  interests}

The authors declare that they have no financial and non-financial conflict of interests about the work reported in this paper.

\section*{Acknowledgements}

This project is supported by the NSFC grant (Nos. 12271334, 12411540016  and 12171306).

\appendix

\section{The Cauchy-Pompeiu integral formula}\label{app-0}

The classical Cauchy-Pompeiu integral formula \cite{P-1909} reads
(see \cite{AF-book-2021}, Lemma 7.6.1)
\begin{eqnarray}\label{Cauchy}
    \psi(p) = \frac{1}{2\pi {\rm i}} \int_{\partial \Omega} \frac{\psi(\mu)}{\mu - p} \,
    \mathrm{d}\mu + \frac{1}{2\pi {\rm i}} \int_{\Omega} \frac{\bar{\partial}\psi(\mu)}{\mu - p} \,
    \mathrm{d}\mu \wedge \mathrm{d}\bar{\mu},
\end{eqnarray}
where $\Omega$ is a finite connected region on the complex plane $\mathbb{C}$, 
and the boundary $\partial \Omega$ is a positively oriented (anticlockwised) simple closed contour.
This formula can be proved when $\psi(p)\in \mathrm{C}^1(\Omega)$,
i.e., if writing  $p=x+\mathrm{i}y$ and $\psi(p)=u(x,y)+\mathrm{i}v(x,y)$, 
the derivatives $u_x,u_y, v_x$ and $v_y$ are continuous for $p\in \Omega$.
However, this formula is also valid when $\bar{\partial}\psi(p)$ is a general function or a distribution
(see, e.g. \cite{BC-IP-1989}).

It is necessary to specify the Dirac $\delta$ function we use in this paper.
Usually the $\delta$ function defined on $\mathbb{C}$ is denoted as
$\delta^{(2)}(p)=\delta(x)\delta(y)$
where $\delta(x)$ and $\delta(y)$ are the usual $\delta$ functions define on the real axis.
This corresponds to the area element setting $\mathrm{d}x\mathrm{d}y$.
In this paper, we adopt the notions in \cite{AF-book-2021}:
we use $\mathrm{d}p \wedge \mathrm{d}\bar{p}$ to represent the area element,
which is related to the former by 
$\mathrm{d}p \wedge \mathrm{d}\bar{p}=-2\mathrm{i}\mathrm{d}x\mathrm{d}y$.
Thus the $\delta$ function used in this paper is $\delta(p) =\frac{\mathrm{i}}{2}\delta^{(2)}(p)$,
from which it follows that \cite{AF-book-2021}
\begin{eqnarray}\label{del}
    \int_\Omega f(p)\delta(p-p_0){\rm d}p\wedge {\rm d}\bar{p}=f(p_0),
\end{eqnarray}
where the region $\Omega$ contains the point $p_0$ and $f(p)$ is 
sufficiently smooth in $\Omega$.

Thus, corresponding the above setting of $\delta(p)$, 
for the operator $\bar{\partial}$, its fundamental solution 
(satisfying $\bar{\partial}F =\delta(p)$)  is \cite{AF-book-2021,H-book-2003}
\begin{eqnarray}
F=\frac{\mathrm{i }}{ 2\pi p}.
\end{eqnarray}
In other words, its Green function of $\bar \partial$ is $\frac{\mathrm{i}}{ 2\pi (\mu-p)}$.
Then, for a Dbar equation 
\begin{equation}\label{fg}
 \bar{\partial}f(p)=g(p)
\end{equation} 
where $f(p)$ vanishes at $\infty$ and $g(p)$ is a distribution, 
its solution can be recovered from  \cite{BC-IP-1989,H-2012},
\begin{eqnarray}\label{fg-Cauchy}
    f(p)=\frac{1}{2\pi {\rm i}} \int_{\mathbb{C}}\frac{g(\mu)}{\mu-p}
    \mathrm{d}\mu \wedge \mathrm{d}\bar\mu.
\end{eqnarray}
Consider a function $\psi(p)$ which admits a ``polynomial normalization'' (see \cite{JM-JMP-1987}) when $p\sim \infty$,
i.e., admitting the following asymptotic behavior at infinity:
\begin{equation}\label{Psi-asy}
\psi(p) \sim \mathrm{Nor}\,\psi(p)+ \sum^{-\infty}_{j=-1} \psi^{(j)} p^j, ~~~(p\sim \infty),
\end{equation}
where (here we employ the short hand used in \cite{JM-JMP-1987,JMM-IP-1988})
\begin{equation}\label{Nor}
\mathrm{Nor}\,\psi(p)=\sum^N_{j=0} \psi^{(j)} p^j, ~~~ (N \geq 0) 
\end{equation}
denotes the polynomial part (the principal  part  plus the constant term in the expansion of $\psi(p)$ at infinity).
Then in \eqref{fg}, letting $f(p)=\psi(p)-\mathrm{Nor}\,\psi(p)$ which vanishes at $\infty$ and 
letting $g(p)=\bar{\partial}f(p)=\bar{\partial}\psi(p)$ which is a distribution,
it follows from \eqref{fg-Cauchy} that
\begin{eqnarray}\label{Cauchy2}
    \psi(p) = \mathrm{Nor}\,\psi(p)
    + \frac{1}{2\pi {\rm i}} \int_{\mathbb{C}} \frac{\bar{\partial}\psi(\mu)}{\mu - p} \,
    \mathrm{d}\mu \wedge \mathrm{d}\bar{\mu}.
\end{eqnarray}
This is an extension of the Cauchy-Pompeiu integral formula \eqref{Cauchy}
where 
\begin{eqnarray}
    \mathrm{Nor}\,\psi(p)=\frac{1}{2\pi {\rm i}} \int_{\partial \mathbb{C}} \frac{\psi(\mu)}{\mu - p}\,
    \mathrm{d}\mu=\underset{\mu=0}{\mathrm{Res}} \left[\frac{\psi(\mu)}{\mu - p}\right].
\end{eqnarray}
Note  here that $\bar{\partial}\psi(\mu)$ is a distribution.

As for the assumption of the uniqueness of solutions of the Dbar problem \eqref{Dp}
where $R(p)$ is a distribution, we have the following remark.

\begin{remark}\label{Rem-3}
Substituting   \eqref{Dp} into \eqref{Cauchy2}
yields 
\begin{eqnarray}\label{CP}
    \psi(p) = \mathrm{Nor}\psi(p)
    + \frac{1}{2\pi {\rm i}} \int_{\mathbb{C}} \frac{\psi(-\mu) R(\mu)}{\mu - p} \,
    \mathrm{d}\mu \wedge \mathrm{d}\bar{\mu}.
\end{eqnarray}
If we only consider those solutions of \eqref{Dp} that are determined through solving the above integral equation,
the assumption that the homogeneous equation \eqref{chi0} admits only zero solution $\psi(p)=0$
is equivalent to the assumption that \eqref{Dp} admits only zero solution $\psi(p)=0$
when $\mathrm{Nor}\,\psi(p)=0$.
Such an assumption was used in the  paper \cite{JM-JMP-1987} and \cite{JMM-IP-1988}.
\end{remark}

\section{Compatibility of Lax pair \eqref{Lax-mkdv2}}\label{app-1}

In this section, we derive \eqref{CCC} from the compatibility of the Lax pair \eqref{Lax-mkdv2}.
First, recall the Lax pair \eqref{Lax-mkdv2}, which can be written as
\begin{subequations}
\begin{align}
& K\varphi(p)=\frac{w}{\alpha^2}\widetilde{\widetilde{\varphi}} (p)
 +\big[(1+w)-\frac{b}{\alpha}(w-1)\big]\widetilde{\varphi}(p)+\alpha^2\varphi(p)=p^2\varphi(p), \label{Lp_1}\\
&\widehat{\varphi}(p)=(BT_n-f_2-c_2)\varphi(p)
= \frac{\beta}{\alpha} s\widetilde{\varphi}(p)+\beta[(\alpha s-\beta)- b(s-1)]\varphi(p),\label{Lp_2}
\end{align}
\end{subequations}
where $K$ is an operator
\begin{subequations}
\begin{equation}\label{K}
 K=\frac{w}{\alpha^2}T_n^2
 +\big[(1+w)-\frac{b}{\alpha}(w-1)\big]T+\alpha^2,
\end{equation}
$T_n$ is the shift operator in $n$-direction, defined in \eqref{TnTm},
and
\begin{align}
& B=\frac{\beta}{\alpha} s,\label{B}\\
& f_2=-\beta(\alpha-b)s, ~\quad c_2=\beta(-b+\beta).
\end{align}
\end{subequations}
To proceed, we decompose the operator $K$ into the following form
\begin{eqnarray}\label{K_2}
    K=(AT_n+f_1+c_1)(BT_n-f_2-c_2)+\sigma,
\end{eqnarray}
where $A, f_1, c_1, \sigma$ are functions to be determined.
Equating the operators \eqref{K}  and \eqref{K_2}, we can obtain
\begin{subequations}
\begin{align}
& A\widetilde{B}=\frac{w}{\alpha^2},\label{A_cc_1}\\
&(1+w)-\frac{b}{\alpha}(w-1)=(f_1+c_1)B
-A(\widetilde{f}_2+c_2), \label{A_cc_2}\\
& (f_1+c_1)(f_2+c_2)-\sigma=-\alpha^2. \label{A_cc_3}
\end{align}
\end{subequations}
Substituting \eqref{B} into \eqref{A_cc_1} we have
\begin{eqnarray}\label{A}
    A=\frac{w}{\alpha \beta \wt s}.
\end{eqnarray}
Due to $K\varphi(p)=p^2\varphi(p)$, which means $K\varphi(p)$ is also a solution of Eq.\eqref{Lp_2},
we have
\begin{align}
 \widehat{K\varphi(p)}&= (BT_n-f_2-c_2)K\psi(p) \nn \\
 &= (BT_n-f_2-c_2)[(AT_n+f_1+c_1)(BT_n-f_2-c_2)+\sigma]\varphi(p) \nn \\
 &= (BT_n-f_2-c_2)(AT_n+f_1+c_1)\widehat{\varphi}(p)+\sigma\widehat{\varphi}(p).
\end{align}
Substituting Eq.\eqref{Lp_1} into the left-hand side of the above equation yields
\begin{subequations}
\begin{align}
 &   B\widetilde{A}=\frac{\widehat{w}}{\alpha^2},\label{A_cc_4} \\ &(1+\widehat{w})-\frac{b}{\alpha}(\widehat{w}-1)=(\widetilde{f}_1+c_1)B-A(f_2+c_2).\label{A_cc_5}
\end{align}
\end{subequations}
Then, eliminating ${f}_1$ from Eq.\eqref{A_cc_2} and \eqref{A_cc_5},
and taking into account of   \eqref{A_cc_1} and \eqref{A_cc_4}, we have
 \begin{eqnarray}
     (T_n-1 ) \left(\frac{\beta(\beta-b)A+1+\frac{b}{\alpha}}{B}-(\alpha^2-\alpha b)A\right)=0.
 \end{eqnarray}
This implies that
\begin{eqnarray}\label{ga}
    \frac{\beta(\beta-b)A+1+\frac{b}{\alpha}}{B}-(\alpha^2-\alpha b)A=\gamma,
\end{eqnarray}
where $\gamma$ is considered as an ``integration'' constant independent of  $n$.
Substituting \eqref{A} and \eqref{B} into \eqref{ga}, we have
\begin{eqnarray}\label{ga-1}
\gamma=\frac{(\beta-b)w}{\beta s\widetilde{s}}+\frac{\alpha+b}{\beta s}-\frac{(\alpha-b)w}{\beta \widetilde{s}}.
\end{eqnarray}
Then, $\gamma$ can be determined from the asymptotics of $s$ and $w$.
For example, we can assume\footnote{
In fact, from \eqref{h-m} and \eqref{g-m} we know that $s$ and $w$ are defined via $\phi^{(-1)}$
which can be formulated in \eqref{phi-l} by
$\phi ^{(-1)}=-\br^T   (\bI+\bM)^{-1}(\bK+b\bI)^{-1}\bc$.
One can always assume
$\alpha>\beta>k_{N}>k_{N-1}> \cdots > k_1 >0$ such that the plane wave factor
$\rho_j$ (see \eqref{pwf-rho}) satisfies $\rho_j\to 0$ when $n\to -\infty$ 
and $\rho_j\to +\infty$ when $n\to +\infty$ for
$j=1,2,\cdots, N$.
This leads to $\phi^{(-1)} \to 0$ when $n\to -\infty$ and $\phi^{(-1)} \to $ constant when $n\to -\infty$,
exhibiting a kink shape.
Then \eqref{sw} follows.
\begin{equation}\label{sw}
s \sim 1,~~ w\sim 1,~~~~ ( n \sim \pm \infty).
\end{equation}
Thus, taking $n\to -\infty$ in \eqref{ga-1} we have
\begin{eqnarray}
    \gamma=1+\frac{b}{\beta}.
\end{eqnarray}
}
With this choice of $\gamma$, Eq.\eqref{ga}  is
\begin{eqnarray}
    \widetilde{s}\big[-b-\alpha+(\beta+b)s \big]=w\big[\beta-b-(\alpha-b)s\big],
\end{eqnarray}
which is \eqref{CC1}.
In addition, eliminating $A$ from \eqref{A_cc_1} and \eqref{A_cc_4},  and taking into account of \eqref{B},
we arrive at
\begin{eqnarray}
     \frac{\widetilde{\widetilde{s}}}{s}=\frac{\widetilde{w}}{\widehat{w}},
  \end{eqnarray}
which is \eqref{CC2}.

\section{A second formulation of the NQC equation}\label{app-2}

\subsection{A second formulation}\label{app-2-1}

There is a second formulation of the NQC equation \eqref{NQCeq}
from the eigenfunction of the Lax pair of the lpKdV equation.
Suppose that $\psi(a)$ and $\psi(b)$ are two eigenfunctions of the lpKdV Lax pair \eqref{Lax-LM}
corresponding to $p=a, b$ and same potential function $h$.
Then, it can be proved that
\begin{equation}\label{Sab-psi}
S(a,b)=\frac{1}{a+b}-\frac{1}{a^2-b^2}\frac{\psi(b)}{\psi(-a)}
\end{equation}
provides a solution to the NQC equation \eqref{NQCeq} \cite{ZZZ-SHU-2022}.

To see such a formulation, let us first  recall the Lax pair \eqref{Lax-LM} of the lpKdV equation, i.e.
\begin{subequations}
\begin{align}
 &     ( \alpha+p ) \widetilde{\widetilde{\psi}} (p)+h\widetilde{\psi}(p)
 + ( \alpha-p )\psi(p)=0,\label{B_L1}\\
& (\beta+p)\widehat{\psi}(p)-(\alpha+p)\widetilde{\psi}(p)+g\psi(p)=0,\label{B_L2}
\end{align}
\end{subequations}
from which we can also obtain
\begin{equation}
 ( \beta+p) \widehat{\widehat{\phi}} (p)+(h+\widetilde{g}+\widehat{g})\widehat{\phi}(p) +( \beta-p)\phi(p)=0,
 \label{B_L3}
\end{equation}
We choose $p=a,b$ in Eq.\eqref{B_L1}, which yields
\begin{subequations}
\begin{align}
& ( \alpha+a ) \widetilde{\widetilde{\psi}} (a)+h\widetilde{\psi}(a)
 + ( \alpha-a )\psi(a) = 0,\\
& ( \alpha+b ) \widetilde{\widetilde{\psi}} (b)+h\widetilde{\psi}(b)
 + ( \alpha-b )\psi(b) = 0.
\end{align}
\end{subequations}
Eliminating $h$ gives rise to
\begin{eqnarray}\label{B.5}
    (\alpha+a)\widetilde{\widetilde{\psi}}(a)\widetilde{\psi}(b)-(\alpha+b)\widetilde{\psi}(a)
    \widetilde{\widetilde{\psi}}(b)-(\alpha-b)\widetilde{\psi}(a)\psi(b)+(\alpha-a)\psi(a)\widetilde{\psi}(b)=0.
\end{eqnarray}
Introduce
\begin{eqnarray}
    X(a,b)=\frac{\psi(b)}{\psi(a)},
\end{eqnarray}
in term of which \eqref{B.5} is rewritten as
\begin{eqnarray}
    (\alpha+a)\widetilde{X}(a,b)-(\alpha+b)\widetilde{\widetilde{X}}(a,b)
    -(\alpha-b)X(a,b)\frac{\psi(a)}{\widetilde{\widetilde{\psi}}(a)}
    +(\alpha-a)\widetilde{X}(a,b)\frac{\psi(a)}{\widetilde{\widetilde{\psi}}(a)}=0.
\end{eqnarray}
This leads to the following relation between $X(a,b)$ and $\psi(a)$:
\begin{eqnarray}
    \frac{(\alpha+b)\widetilde{\widetilde{X}}(a,b)-(\alpha+a)\widetilde{X}(a,b)}{(\alpha-a)\widetilde{X}(a,b)
    -(\alpha-b)X(a,b)}=\frac{\psi(a)}{\widetilde{\widetilde{\psi}}(a)}.
\end{eqnarray}
Along the same line, from Eq.\eqref{B_L2} and \eqref{B_L3} we can have
\begin{subequations}
\begin{align}
&    \frac{(\beta+b)\widehat{X}(a,b)-(\beta+a)X(a,b)}{(\alpha+b)\widetilde{X}(a,b)-(\alpha+a)X(a,b)}
=\frac{\widetilde{\psi}(a)}{\widehat{\psi}(a)},
\\
&    \frac{(\beta+b)\widehat{\widehat{X}}(a,b)-(\beta+a)\widehat{X}(a,b)}{(\beta-a)\widehat{X}(a,b)-(\beta-b)X(a,b)}
=\frac{\psi(a)}{\widehat{\widehat{\psi}}(a)}.
\end{align}
\end{subequations}
These three equations, in light of the equality
\begin{eqnarray}
\frac{\psi(a)}{\widetilde{\widetilde{\psi}}(a)}.\widetilde{\bigg(\frac{\widetilde{\psi}(a)}{\widehat{\psi}(a)}\bigg)}.
\widehat{\bigg(\frac{\widetilde{\psi}(a)}{\widehat{\psi}(a)}\bigg)}=\frac{\psi(a)}{\widehat{\widehat{\psi}}(a)},
\end{eqnarray}
yields an equation for $X(a,b)$:
\begin{eqnarray}\label{Xab}
    \frac{(\beta+b)\widetilde{\widehat{X}}(a,b)-(\beta+a)\widetilde{X}(a,b)}{(\alpha-a)\widetilde{X}(a,b)
    -(\alpha-b)X(a,b)}.\frac{(\beta-a)\widehat{X}(a,b)-(\beta-b)X(a,b)}{(\alpha+b)\widetilde{\widehat{X}}(a,b)
    -(\alpha+a)\widehat{X}(a,b)}=1.
\end{eqnarray}
Now we introduce $S(a,b)$ by
\begin{eqnarray}\label{X-S}
    X(-a,b)=(b-a)\big[1-(b+a)S (a,b)\big]
\end{eqnarray}
which yields \eqref{Sab-psi},
and it follows from \eqref{Xab} that
\begin{eqnarray}
    \frac{1-(\beta+b)\widetilde{\widehat{S}} (a,b)+(\beta-a)\widetilde{S} (a,b)}
    {1-(\alpha+a)\widetilde{S} (a,b)+(\alpha-b)S (a,b)}.
    \frac{1-(\beta+a)\widehat{S} (a,b)+(\beta-b)S (a,b)}
    {1-(\alpha+b)\widetilde{\widehat{S}} (a,b)+(\alpha-a)\widehat{S} (a,b)}=1,
\end{eqnarray}
which is the NQC equation \eqref{NQCeq}.

\subsection{Comparison of different formulations }\label{app-2-2}

The above formulation \eqref{Sab-psi} is different from ours.
We explain the difference in what follows.
First, in Sec.\ref{sec-4} explicit formulae of the functions in the Lax pair have been derived.
From these expressions we know that the involved functions, namely, $\phi(p)$, $w$ and $s$,
allow expansions in terms of $b$ at infinity.
For $\phi(p)$, from \eqref{phi-p} we have
\begin{equation}\label{phi-pp}
\phi(p)=\phi^{[-1]} b^{-1}+ \phi^{[-2]} b^{-2}+ \cdots,
\end{equation}
where
\begin{equation}\label{Vp}
\phi^{[-1]}=1-\br^T(\bK+p \bI)^{-1}(\bI+\bM)^{-1}\bc=V(p).
\end{equation}
Note here that we use the superscript $[-j]$ in $\phi^{[-j]}$ so that we distinguish them with $\phi^{(-j)}$
in Eq.\eqref{Laurent-phi}.
We also connect $\phi^{[-1]}$ with the Cauchy matrix variable $V(p)$ (see \eqref{Va}).
In addition, for $\phi^{(-1)}$ given in \eqref{phi-l}, we have
\begin{equation}\label{phi-ll}
\phi^{(-1)}=\phi^{[-1,-1]} b^{-1}+ \phi^{[-1,-2]} b^{-2}+ \cdots,
\end{equation}
where
\begin{equation}
\phi^{[-1,-1]}=-\br^T \left(\bI+\bM\right)^{-1}\bc=-S^{(0,0)}.
\end{equation}
Substituting \eqref{phi-ll} into \eqref{h-m} and \eqref{g-m}, we get
\begin{equation}\label{w-exp}
w=1+\left( \phi^{[-1,-1]}-\wt{\wt{\phi}}{}^{[-1,-1]}\right) b^{-1} + O(b^{-2})
\end{equation}
and
\begin{equation}\label{s-exp}
s=1+\left( \widehat{\phi}^{[-1,-1]}-\wt{\phi}^{[-1,-1]}\right) b^{-1} + O(b^{-2}).
\end{equation}
Now we substitute \eqref{phi-pp}, \eqref{w-exp} and \eqref{s-exp} into the Lax pair \eqref{Lax-mkdv}.
Its leading term (which is $b^{-1}$-term) gives rise to the lpKdV Lax pair \eqref{Lax-LM}
with correspondence
\begin{align}
& \psi(p)=\phi^{[-1]}=V(p),\\
& h=-2\alpha - \phi^{[-1,-1]}+\wt{\wt{\phi}}{}^{[-1,-1]}=-2 \alpha +S^{(0,0)}-\wt{\wt{S}}{}^{(0,0)},\\
& g=\alpha-\beta +\widehat{\phi}^{[-1,-1]}-\wt{\phi}^{[-1,-1]}= \alpha-\beta -\wh{S}^{(0,0)}+\wt{S}^{(0,0)}.
\end{align}
Thus, we have recovered the lpKdV Lax pair \eqref{Lax-LM} form the expansion of \eqref{Lax-mkdv}.
In light of the above correspondence, the formulation \eqref{Sab-psi} indicates
\begin{equation}\label{B.23}
S(a,b)=\frac{1}{a+b}-\frac{1}{a^2-b^2}\frac{V(b)}{V(-a)}
\end{equation}
where $V(p)$ is formulated by
\begin{eqnarray}
    V(p)=1-\br^T(\bK+p \bI)^{-1}(\bI+\bM)^{-1}\bc.
\end{eqnarray}
The above $S(a,b)$ is apparently different from our formulation \eqref{Sab}, i.e. \eqref{S-phi},
because \eqref{B.23} does not admit the symmetry property $S(a,b)=S(b,a)$
but  \eqref{Sab} does.
In fact, if one assumes  $S(a,b)=S(b,a)$, 
then it must follow that $-V(b)V(-b)=V(a)V(-a)$, 
which   obviously does not hold  when $a,b \to \infty$ 
because $V(p)\to 1$ as $p\to \infty$.
The simplest solution of this case ($N=1$) is
\begin{eqnarray}
S(a,b)= \frac{1}{a+b} - \frac{(a - k_1)(b-k_1) (2 k_1 + \rho_1)+4k_1^2 (a - k_1)}
{(a^2 - b^2)[(a+k_1)(b + k_1)(2 k_1 + \rho_1)-4k^2_1(b+k_1)]},
\end{eqnarray}
where $ \rho_1$ is defined in \eqref{pwf-rho}.

We have shown that the NQC variable $S(a,b)$ we introduced in Sec.\ref{sec-3-3}
has an explicit form \eqref{Sab}, i.e.
\begin{eqnarray}\label{Sab-app-C}
    S(a,b)=\br^T (\bK+a\bI)^{-1} \left(\bI+\bM\right)^{-1}(\bK+b\bI)^{-1}\bc,
\end{eqnarray}
which is the same as the one formulated in the Cauchy matrix approach, i.e. \eqref{Sab-1}.
Note that the symmetry property $ S(a,b)= S(b,a)$ for the above expression has been proved in \cite{ZZ-SAPM-2013}.

In the following, let us identify the $S(a,b)$ defined in \cite{SZZ-arXiv-2025},
which follows from  \eqref{5.38} that 
\begin{eqnarray}\label{5.38-C}
 S(a,b)=\frac{1}{a+b}-\frac{1}{b^2-a^2} D(a,b).
\end{eqnarray}
Note that the definition \eqref{D-pq} for $D(p,q)$ indicates that $D(a,b)=-D(b,a)$,
which gives rise to the symmetry property $S(a,b)= S(b,a)$ as well.
Then, in light of the correspondence \eqref{2.32d}, the above $S(a,b)$ in \eqref{5.38-C} can be written as
\begin{eqnarray}
    S(a,b)=\frac{1}{a+b}-\frac{1}{b^2-a^2}[(\alpha+b)V(a)\widetilde{V}(b)-(\alpha+a)\widetilde{V}(a)V(b)],
\end{eqnarray}
where $V(a)$ is defined as \eqref{Va}.
This is actually the same as the alternative expression \eqref{Sdbv} for $S(a,b)$
which is constructed in the Cauchy matrix approach.
Thus, we can conclude that  the NQC variable $S(a,b)$ formulated in Sec.\ref{sec-3} of the present paper,
formulated in the Cauchy matrix approach and formulated in Ref.\cite{SZZ-arXiv-2025} are same.
The simplest one ($N=1$) reads 
\begin{eqnarray}
     S(a,b)=\frac{2k_1\rho_1}{(k_1+a)(k_1+b)(2k_1+\rho_1)}
\end{eqnarray}
where $ \rho_1$ is defined in \eqref{pwf-rho}.

\end{document}